\documentclass[%
 reprint,
 amsmath,amssymb,
 aps,
]{revtex4-2}

\usepackage{graphicx}
\usepackage{dcolumn}
\usepackage{bm}
\usepackage{lipsum}
\usepackage{color,soul}
\usepackage{dsfont}
\usepackage[colorinlistoftodos]{todonotes}
\usepackage{booktabs}

\usepackage{makecell}

\begin{document}

\preprint{APS/123-QED}

\title{Inter-band B(E1) Strengths in  Heavy Nuclei \\based on the Proxy-$SU(3)$ Scheme }

\author{A. Restrepo}
 \altaffiliation{alejandro.restrepo19@udea.edu.co}
\author{J. P. Valencia}%
 \email{patricio.valencia@udea.edu.co}
\affiliation{Instituto de Física, Universidad de Antioquia, Calle 70 No. 52-21, Medellín, Colombia}%




\date{\today}

\begin{abstract}

Rotational bands of positive and negative parity are frequently found experimentally in heavy deformed even-even nuclei along with their corresponding $E2$ intra-band and $E1$ inter-band transitions. In $SU(3)$ symmetry-based models, calculating the $B(E2)$ strengths is relatively straightforward due to the relation between the $E2$ operator and the quadrupole generator of the algebra. On the other hand, calculating the $B(E1)$ strengths is a more challenging task, requiring computations involving the shell-model expansion of the dipole operator and multi-shell matrix elements. In this work the $E1$ operator is expanded in $SU(3)$ second quantization formalism and applied to calculate the inter-band $B(E1)$ strengths of the isotope $^{224}$Th using the semi-microscopic algebraic quartet model (SAQM) based on the recently proposed proxy-$SU(3)$ scheme.

\end{abstract}


\maketitle


\section{\label{sec:theory}INTRODUCTION}

The efforts to recover $SU(3)$ symmetry in heavy deformed nuclei have resulted in a few models developed over the years \cite{osti_4174532,RAJU1973433,PhysRevC.52.R1741,PhysRevC.92.024320, doi:10.1142/5385} with good results describing internal structure, deformation parameters and electromagnetic interactions. The most recent proposal is the proxy-$SU(3)$ scheme \cite{PhysRevC.95.064325,sym15010169}, which restores the symmetry approximately using the property of high overlap between intruder and deserter Nilsson orbitals with minimum truncation of the complete space \cite{PhysRevC.88.054309}.

This scheme allows the extension of $SU(3)$ results to heavy nuclei regions so that properties like low energy structure, deformation parameters and electromagnetic transitions can be studied in a symmetry-based scheme. Some applications can be found in the literature \cite{PhysRevC.95.064326,Martinou2017ParameterFP} with good agreement with experimental observations and several predictions for values yet to be measured. In reference \cite{PhysRevC.101.054306}, the semi-microscopic algebraic quartet model (SAQM) \cite{CSEH2015213} in the proxy-$SU(3)$ scheme is applied to the actinide $^{224}$Th, where calculations for deformation parameters $(\beta,\gamma)$, intra-band $B(E2)$ strengths, low energy excitation spectra and band heads are shown, providing valuable insight into the capabilities of the model.

It is the intention of this article to extend the results of \cite{PhysRevC.101.054306} for the isotope $^{224}$Th by computing the expansion of the $E1$ transition operator in $SU(3)$ second quantization for the actinide region and obtain the inter-band $B(E1)$ strengths using the irreducible representations (irreps) of the SAQM in the proxy-$SU(3)$ scheme. This result continues the theoretical modeling of the observed low-energy structure of such isotope and allows analogous extensions to other nuclei. 

Even though, experimental values of $B(E1)$ strengths for $^{224}$Th have not been reported yet, the results are compared with available ratios $R(E1/E2) = B(E1)/B(E2)$ obtained from static dipole to quadrupole $D_0/Q_0$ measures reported in \cite{ACKERMANN199361, SCHULER1986241, LEANDER198658}. The absence of experimental
data for this particular nucleus is challenging to under-
standing its internal structure, we hope to stimulate
research towards this direction in future experiments

An additional motivation for studying the positive-negative parity band structures and their associated inter-band transitions strengths observed in many heavy even-even nuclei is their relation to static parity-breaking octupole moments \cite{NAZAREWICZ1990c333, LEANDER198658, SCHULER1986241,OctupoleButler}, which are a rich source of novel nuclear theory. This article begins explaining the model space for ground and excited states in the proxy-$SU(3)$ scheme, followed by the expansion of the $E1$ operator and computation of its matrix elements. These calculations are subsequently applied to study $^{224}$Th where its results are compared to available data. 

\section{MODEL SPACE}

In heavy nuclei, protons and neutrons occupy different valence shells, so their permutational symmetries must be considered independently. In the SAQM, the ground state ($0\hbar\omega$) spatial representations for protons ($\pi$) and neutrons ($\nu$) are $[f]_\pi=[2^{I}]_\pi$ and $[f]_\nu=[2^{I}]_\nu$ respectively, where $I$ is the number of quartets in each valence shell. These irreps are considered by two reasons; first, they have the highest spatial symmetry which is favored by the dominant quadrupole residual interaction and second, they can couple to the quartet symmetry defined as $[4^I]$.

The next step in defining the nuclear quantum state is the decomposition of the permutational irreps into irreps of a trivial algebra. Since the symmetry of the model is based on the  $SU(3)$ algebra, we are particularly interested in obtaining its quantum numbers by solving the branching problem $U(\Omega)\supset SU(3)$ 
\cite{DRAAYER1989279}. This decomposition must be done independently for protons and neutrons, expressed as $U(\Omega_\pi)\supset SU_{\pi}(3)$ and $U(\Omega_{\nu})\supset SU_{\nu}(3)$ respectively. Then, the model space is truncated to the leading irrep of each $SU(3)_\pi$ and $SU(3)_\nu$ since they contribute the most to the residual interaction. The total irrep is obtained as the leading $(\lambda,\mu)$ contained in the product $(\lambda_\pi,\mu_\pi)\otimes(\lambda_\nu,\mu_\nu)$.

The $L-S$ coupling is adopted in the subsequent branching algebras in order to exploit the spatial symmetry of the system via the physical basis $SU(3)\supset SO(3)\supset SO(2)$.  All allowed orbital angular momentum values $L$ will be those contained in the irrep $(\lambda,\mu)$ along with their corresponding multiplicities $K$. The spin labels are obtained from the conjugate representations $[\overline{2^{I}}]_\pi=[I^{2}]_\pi$ and $[\overline{2^{I}}]_\nu=[I^{2}]_\nu$ where their coupling results in total $S = 0$. The complete algebra chain and corresponding ground state band $|\Psi_{g.s.b.}(J^{\Pi})\rangle$ are shown in equations \ref{pn_groundchain} and \ref{pn_groundstate1} respectively, where the spin labels are tacit.

\begin{widetext}

\begin{equation}
\begin{aligned} 
U\left(2\Omega_{\pi}\right)\otimes U\left(2\Omega_{\nu}\right)&\supset\Bigg\{U\left(\Omega_{\pi}\right)\otimes U\left(\Omega_{\nu}\right)\supset SU_{\pi}(3)\otimes SU_{\nu}(3)\supset SU_{\pi+\nu}(3)\supset SO_{\pi+\nu}(3)\Bigg\}\\&\otimes\Bigg\{ SU_{\pi}(2)\otimes SU_{\nu}(2)\supset SU_{\pi+\nu}(2) \Bigg\}\supset SU_{J}(2)\supset U_{M}(1),
\end{aligned} 
\label{pn_groundchain}
\end{equation}


\begin{equation}
\begin{aligned} 
&|\Psi_{g.s.b.}(J^{\Pi})\rangle=\Bigg|\hspace{1
mm}[2^{I}]_{\pi}\otimes[2^{I}]_{\nu},\hspace{1mm}(\lambda_{\pi},\mu_{\pi})\otimes(\lambda_{\nu},\mu_{\nu}),\hspace{1mm}\rho(\lambda,\mu),\hspace{0.4mm} K,\hspace{0.4mm} L;\hspace{1mm}J = L, M\Bigg>,
\end{aligned} 
\label{pn_groundstate1}
\end{equation}

\begin{equation}
\resizebox{0.95\hsize}{!}{$
\begin{aligned} 
U\left(2(\Omega_{\pi-}+\Omega_{\pi+})\right)\otimes U\left(2\Omega_{\nu}\right)&\supset\Bigg\{ \Big(U\left(\Omega_{\pi-}\right)\otimes U\left(\Omega_{\pi+}\right)\Big)\otimes U\left(\Omega_{\nu}\right)\supset \Big(SU_{\pi-}(3)\otimes SU_{\pi+}(3)\Big)\otimes SU_{\nu}(3)\supset  SU_{\pi}(3)\otimes SU_{\nu}(3)\\&\hspace{3mm}\supset SU_{\pi+\nu}(3)\supset SO_{\pi+\nu}(3)\Bigg\}\\&\otimes\Bigg\{ \Big(SU_{\pi-}(2)\otimes SU_{\pi+}(2)\Big)\otimes SU_{\nu}(2)\supset SU_{\pi}(2)\otimes SU_{\nu}(2)\supset SU_{\pi+\nu}(2) \Bigg\}\supset SU_{J}(2)\supset U_{M}(1),
\end{aligned} 
\label{pn_excitedchain}
$}
\end{equation}

\begin{equation}
\resizebox{0.95\hsize}{!}{$
\begin{aligned} 
|\Psi_{e.b.}^{\pi}(J^{\Pi})\rangle=\Bigg|\hspace{1mm}\Big([2^{I-1},1]_{\pi}\otimes[1]_{\pi}\Big)\otimes[2^{I}]_{\nu},\hspace{1mm}\Big((\lambda_{\pi-},\mu_{\pi-})\otimes(\lambda_{\pi+},\mu_{\pi+})\Big)\otimes(\lambda_{\nu},\mu_{\nu}),\hspace{1mm}(\lambda_{\pi},\mu_{\pi})\otimes (\lambda_{\nu},\mu_{\nu}),\hspace{1mm}\rho(\lambda,\mu)&,K,L;\hspace{1mm}J=L,M\Bigg>.
\end{aligned} 
$}
\label{pn_excitedstate1}
\end{equation}

\end{widetext}




The $1\hbar\omega$ excitation mechanism of the model has two contributions; the independent one proton and one neutron excitations to their corresponding contiguous upper shells. Thus, the model space of the excited nucleon becomes a  two-shell system \cite{Isacker_2016} and its symmetry algebra turns into a product of two algebras with irreps $[2^{I-1},1]\otimes[1]$, for the lower and upper shell respectively. These are treated independently in their branching, resulting in a product $SU(3)\otimes SU(3)$ which is truncated to the leading irreps of each factor. This product is finally coupled to a total $SU(3)$ and truncated again. The non-excited nucleon space remains just like in the ground state case.

Take as an example the proton excitation mechanism: the product between lower and upper proton shells leading irreps with neutron leading irrep is denoted as   $(\lambda_{\pi_-},\mu_{\pi_-})\otimes(\lambda_{\pi_+},\mu_{\pi_+})\otimes(\lambda_\nu,\mu_\nu)$, which couples to leading $(\lambda_{\pi},\mu_{\pi})\otimes(\lambda_\nu,\mu_\nu)$ and then to the total leading $(\lambda,\mu)$. The  orbital angular momentum and spin labels are obtained in the same way as in the ground state case but only couplings to total $S=0$ are considered. The proton excitation algebra chain and corresponding excited band state $|\Psi_{e.b.}^{\pi}(J^{\Pi})\rangle$ are shown in equations \ref{pn_excitedchain} and \ref{pn_excitedstate1} respectively where the spin labels are tacit. The neutron excitation scheme is analogous and couples to the same leading irrep $(\lambda,\mu)$.

\section{$E1$ TRANSITION OPERATOR}

The electric dipole transition operator in the long wavelength approximation is defined as \cite{brussaard1977shell}
\begin{equation}
\begin{aligned} 
&T^{1}_{\mu}(E)= b_0\Big(e_{\pi}\mathcal{Q}^{\hspace{2mm}1}_{\pi\mu}+ e_{\nu}\mathcal{Q}^{\hspace{2mm}1}_{\nu\mu}\Big),\\&\mathcal{Q}^{\hspace{2mm}1}_{\sigma\mu}=\sum_{i=1}^{Z(N)}r_{\sigma}(i)Y^{1}_{\mu}(\hat{r}_\sigma(i)),
\end{aligned} 
\label{operator}
\end{equation}
where $\sigma=\pi,\nu$, the corresponding proton and neutron charges are $e_{\pi}$, $e_{\nu}$ and the harmonic oscillator amplitude parameter is $b_0 = A^{1/6}$ fm. The charges $e_{\pi}$ and $e_{\nu}$ are different from the actual $e'_{\pi} = +1e$ and $e'_{\nu} = 0e$, they are effective charges that encompass nuclear core dynamics not explicitly included in the  model, like polarization and deformation due to single-nucleon excitation. For dipole transitions, the value of $e_{\nu}$ is negative \cite{Eisenberg1976}.

The reduced transition probability $B(E\lambda)$ expresses the likelihood of a nucleus undergoing a particular transition and its corresponding decay constant and half-life can be derived from it. For the dipole operator $T^1_\mu(E)$, it is given by
\begin{equation}
\resizebox{0.9\hsize}{!}{$
\begin{aligned} 
B(E1;\xi J\rightarrow\xi' J')=\frac{2J'+1}{2J+1}|\langle\xi' J'||\hspace{1mm}T^{1}(E)\hspace{1mm}||\xi J\rangle|^2,
\end{aligned} 
\label{BE1}
$}
\end{equation}
where $J$ ($J'$) is the total angular momentum of the initial (final) state and $\xi$ ($\xi'$) represents the other algebras labels which in our case of interest will be $\xi = [f],(\lambda,\mu),L,[\bar{f}],S$.

The spherical tensor operator $\mathcal{Q}^{\hspace{2mm}1}_{\sigma\mu}$ has to be expanded in terms of $SU(3)$ tensors in order to be applied  over the model space states $|\Psi(J^{\Pi})\rangle$. This can be done by performing the ordinary expansion in the second quantization operators $a^{\dagger}_{(\eta',0),l',m_{l'},\frac{1}{2},m_{s'}}$, $\tilde{a}_{(0,\eta)l,m_l,\frac{1}{2},m_s}$, which after coupling all quantum numbers forms a sum of irreducible tensors labeled by $\rho_o,(\lambda_o,\mu_o),K_o,L_o,S_o,J_o,M_o$ as explained in appendix A. The result of this calculation is shown in equation \ref{operatorE1}, where the coefficients $\mathcal{V}^{L_o=1}_{\sigma}(\eta',\eta,\lambda_o,\mu_o,K_o)$ contain the products of $SU_J(2)$ reduced matrix elements \cite{moshinsky1996harmonic} and coupling coefficients. Its values are shown in table \ref{table1} for the actinide region where the valence shells are $\eta = 5$ for protons and $\eta = 6$ for neutrons. Notice that no sum over $K_o$ appears since $L_o=1$ states have no multiplicity in a given irrep $(\lambda,\mu)$.

\begin{widetext}
\begin{equation}
\resizebox{0.82\hsize}{!}{$
\begin{aligned} 
\mathcal{Q}^{\hspace{2mm}1}_{\sigma\mu} = &\sum_{(\lambda_o,\mu_o)}\sum_{l,l'}(-1)^{\eta}\sqrt{\frac{2(2l'+1)}{3}}\Big\langle \eta',l',\frac{1}{2} \Big|\Big| r_{\sigma}Y^{1}(\hat{r}_{\sigma}) \Big|\Big| \eta,l,\frac{1}{2}\Big\rangle\\&\times\langle (\eta',0),0,l'; (0,\eta),0,l||(\lambda_o,\mu_o),K_o,L_o=1 \rangle_{\rho_o=1}\Big\{a^{\dagger}_{(\eta',0)\frac{1}{2}}\times \tilde{a}_{(0,\eta)\frac{1}{2}}\Big\}^{\rho_o=1(\lambda_o,\mu_o)K_oL_o=J_o=1}_{\hspace{25mm}M_{J_o}}\\&=\sum_{(\lambda_o,\mu_o)}\mathcal{V}^{L_o=1}_{\sigma}(\eta',\eta,\lambda_o,\mu_o,K_o)\Big\{a^{\dagger}_{(\eta',0)\frac{1}{2}}\times \tilde{a}_{(0,\eta)\frac{1}{2}}\Big\}^{\rho_o=1(\lambda_o,\mu_o)K_oL_o=J_o=1}_{\hspace{25mm}M_{J_o}},
\end{aligned} 
\label{operatorE1}
$}
\end{equation}
\end{widetext}

\begin{table*}[!t]
\caption{\label{table1}
$\mathcal{V}_{\sigma}^{L_o =1}(\eta,\eta',\lambda_o,\mu_o, K_o)$ coefficients of $\mathcal{Q}^{\hspace{2mm}1}_{\sigma\mu}$ tensor expansion for the actinide region.} 
\begin{ruledtabular}
\begin{tabular}{cccccccc}
& \multicolumn{7}{c}{$(\lambda_o,\mu_o)$}\\ 
\cmidrule(l){2-8}
&(6,7)&(5,6)&(4,5)&(3,4)&(2,3)&(1,2)&(0,1)\\
\hline
&&&&&&\\
$\mathcal{V}_{\pi}^{1}(5,6,\lambda_o,\mu_o, K_o)$&   & -0.5245 & 0.2910 & 0.7517
& -0.6500 & -2.4657 & 2.4376 \\
$\mathcal{V}_{\nu}^{1}(6,7,\lambda_o,\mu_o, K_o)$& -0.4823 & -0.3451 & 0.5421 & 0.7759 & -0.9169 & -2.9365 & 3.0565 \\
\end{tabular}
\end{ruledtabular}
\end{table*}

\begin{widetext}
\begin{equation}
\resizebox{0.95\hsize}{!}{$
\begin{aligned} 
\Big\langle \Psi'_{g.s.b.}(J'^{\Pi'})\Big|\Big|&
T^{1}(E)
\Big|\Big|\Phi_{e.b.}(J^{\Pi})\Big\rangle=\\\\&b_0
\Biggl[e_{\pi}\sqrt{\alpha}\sum_{(\lambda_{\pi_o},\mu_{\pi_o})}\mathcal{V}_{\pi}^{1}(\eta',\eta,\lambda_{\pi_o},\mu_{\pi_o}, k_{\pi_o})\sum_{ \bar{\rho},\rho_{\pi}}\langle (\lambda,\mu),K,L; (\lambda_{\pi_o},\mu_{\pi_o}),L_{\pi_o}=1||(\lambda',\mu'),K',L' \rangle_{\bar{\rho}}\begin{Bmatrix}
(\lambda_{\pi},\mu_{\pi}) & (\lambda_{\pi_o},\mu_{\pi_o}) &  (\lambda'_{\pi},\mu'_{\pi}) &  \rho_{\pi} \\
(\lambda_{\nu},\mu_{\nu}) & (0,0) &  (\lambda_{\nu},\mu_{\nu}) &  1 \\
(\lambda,\mu) & (\lambda_{\pi_o},\mu_{\pi_o}) &  (\lambda',\mu') &  \bar{\rho} \\
\rho & 1 & \rho' &  \\
\end{Bmatrix}
\\\\&  \hspace{20mm}\times\Big\langle N'_{\pi},\hspace{1mm} [f'_{\pi}],\hspace{1mm}(\lambda'_{\pi},\mu'_{\pi})\hspace{1mm}\Big|\Big|\Big|  \Big\{a^{\dagger}_{(\eta',0)\frac{1}{2}}\times \tilde{a}_{(0,\eta)\frac{1}{2}}\Big\}^{(\lambda_{\pi_o},\mu_{\pi_o})}\Big|\Big|\Big| N_{\pi},\hspace{1mm} [f_{\pi-}]\otimes[f_{\pi+}],\hspace{1mm}(\lambda_{\pi-},\mu_{\pi-})\otimes(\lambda_{\pi+},\mu_{\pi+}),\hspace{1mm}(\lambda_{\pi},\mu_{\pi})\Big\rangle_{\rho_{\pi}}
\\\\&+e_{\nu}\sqrt{1-\alpha}\sum_{(\lambda_{\nu_o},\mu_{\nu_o})}\mathcal{V}_{\nu}^{1}(\eta'',\eta''',\lambda_{\nu_o},\mu_{\nu_o}, k_{\pi_o})\sum_{\bar{\rho},\rho_{\nu}}\langle (\lambda,\mu),K,L; (\lambda_{\nu_o},\mu_{\nu_o}),L_{\nu_o}=1||(\lambda',\mu'),K',L' \rangle_{\bar{\rho}}\begin{Bmatrix}
(\lambda_{\pi},\mu_{\pi}) & (0,0) &  (\lambda_{\pi},\mu_{\pi}) &  1 \\
(\lambda_{\nu},\mu_{\nu}) & (\lambda_{\nu_o},\mu_{\nu_o}) &  (\lambda'_{\nu},\mu'_{\nu}) &  \rho_{\nu} \\
(\lambda,\mu) & (\lambda_{\nu_o},\mu_{\nu_o}) &  (\lambda',\mu') &  \bar{\rho} \\
\rho & 1 &  \rho' &  \\
\end{Bmatrix}
\\\\&  \hspace{20mm}\times\Big\langle N'_{\nu},\hspace{1mm} [f'_{\nu}],\hspace{1mm}(\lambda'_{\nu},\mu'_{\nu})\hspace{1mm}\Big|\Big|\Big|  \Big\{a^{\dagger}_{(\eta'',0)\frac{1}{2}}\times \tilde{a}_{(0,\eta''')\frac{1}{2}}\Big\}^{(\lambda_{\nu_o},\mu_{\nu_o})}\Big|\Big|\Big| N_{\nu},\hspace{1mm} [f_{\nu-}]\otimes[f_{\nu+}],\hspace{1mm}(\lambda_{\nu-},\mu_{\nu-})\otimes(\lambda_{\nu+},\mu_{\nu+}),\hspace{1mm}(\lambda_{\nu},\mu_{\nu})\Big\rangle_{\rho_{\nu}}\Biggr].
\end{aligned} 
\label{Reducedmatrixelement}
$}
\end{equation}
\end{widetext}

This expansion allows for the calculation of $SU_J
(2)$ reduced matrix elements required for computing $B(E1)$ strengths of equation \ref{BE1}. These matrix elements must take into account the proton and neutron excitations independently according to the model space formulated in equations \ref{pn_groundchain}-\ref{pn_excitedstate1}. Thus it requires a recoupling of its quantum numbers in order to obtain a separate form \cite{TROLTENIER199553, RAJU1973433, CASTANOS1987290, Millener1991SU3IS} as is shown in equation \ref{Reducedmatrixelement}.

This equation represents the transitions from the excited band to the ground state band where $|\Phi_{e.b.}(J^{\Pi})\rangle$ includes the excitations from both protons and neutrons to their corresponding upper shells as a normalized superposition $|\Phi_{e.b.}(J^{\Pi})\rangle = \sqrt{\alpha}|\Psi_{e.b.}^{\pi}(J^{\Pi})\rangle + \sqrt{1-\alpha}|\Psi_{e.b.}^{\nu}(J^{\Pi})\rangle$. The parameter $\alpha$ expresses the contribution of each nucleon type  wavefunction to the excited state of the nucleus and should be determined from experimental data. The factors $\{...\}$ are the 9-$(\lambda,\mu)$ symbols and $\langle...;...||...\rangle$ are the reduced $SU(3)$ Wigner coefficients, both of which can be obtained by already developed libraries in  \cite{DYTRYCH2021108137}. The triple-barred matrix elements $\langle...|||...|||...\rangle$ are reduced with respect to $SU(3)$ and its single-shell values can be computed by the libraries in \cite{BAHRI199459, J2007ReducedME}. Since the present case involves the transitions of nucleons between two shells, this reduced matrix element library cannot be applied directly. A further calculation must be performed in order to obtain such matrix elements which is explained in appendix B.

\section{APPLICATION TO $^{224}\text{Th}$}

The isotope $^{224}$Th has theoretical and experimental interest due to its alleged octupole static deformation predicted in \cite{MOLLER20161} to have a value of $\epsilon_3= 0.11$, along with its low energy spectra that suggests it \cite{OctupoleButler}. The SAQM in the proxy-$SU(3)$ scheme was applied to this isotope in \cite{PhysRevC.101.054306} formulating a two parameter model for the low energy structure, one parameter intra-band $B(E2)$ transition strengths and parameter-less deformation $(\beta,\gamma)$ values. The inter-band $B(E1)$ strengths are studied in this section under such model which hopes to provide more insight about the structure and geometric shape of this isotope from the symmetry-based perspective.

Since this isotope has eight protons and eight neutrons in their respective ground state valence shells $\eta=5$ and $\eta = 6$, their permutational symmetries are $[2^4]$ for both $U_{\pi}(21)$ and $U_{\nu}(28)$. The reductions $U_{\pi}(21)\supset SU_{\pi}(3)$ and $U_{\nu}(28)\supset SU(3)$ result in the leading irreps $(26,4)$ and $(34,4)$ respectively, which couple to the leading $(60,8)$ irrep. This representation contains a $K^{\Pi}=0^+$ Elliott label and even angular momentum $L$ values that constitute the ground state band observed experimentally. 

For the one particle excitation model space, the representation for the excited nucleon is $[2^3,1]\otimes [1]$. In the example of a one proton excitation it is $[2^3,1]_{\pi_-}\otimes [1]_{\pi_+}$ for the algebras $U_{\pi_-}(21)\otimes U_{\pi_+}(28)$ and $[2^4]_{\nu}$ for the ground state $U_{\nu}(28)$ algebra. Such irreps decompose into the leading $(25,2)\otimes(6,0)$ and $(34,4)$ respectively. Coupling the proton space to the neutron space results in the leading irrep $(65,6)$. Both mechanisms of proton and neutron excitation couple to leading $(65,6)$. This representation contains a $K^{\Pi}=0^-$ and odd $L$ values which account for the odd parity band observed experimentally. The model space states explained in this paragraph are shown in equations \ref{pn_groundstateTh} and \ref{pn_excitedstateTh}.

\begin{widetext}
\begin{equation}
\begin{aligned} 
&|\Psi^{^{224}\text{Th}}_{g.s.b.}(J^{\Pi=+})\rangle=\Bigg|\hspace{1
mm}[2^{4}]\otimes[2^{4}],\hspace{1mm}(34,4)\otimes(26,4),\hspace{1mm}1(60,8),\hspace{0.4mm} 0,\hspace{0.4mm} L;J = L,M\Bigg>,
\end{aligned} 
\label{pn_groundstateTh}
\end{equation}

\begin{equation}
\resizebox{0.93\hsize}{!}{$
\begin{aligned} 
|\Psi^{\pi,^{224}\text{Th}}_{e.b.}(J^{\Pi=-})\rangle=\Bigg|\hspace{1mm} &\Big([2^3,1]\otimes[1]\Big)\otimes[2^4],\hspace{1mm}\Big((25,2)\otimes(6,0)\Big)\otimes(34,4),\hspace{1mm}(31,2)\otimes (34,4),\hspace{1mm}1(65,6),\hspace{0.4mm} 0,\hspace{0.4mm} L ;J=L,M\Bigg>.
\end{aligned} 
\label{pn_excitedstateTh}
$}
\end{equation}
\end{widetext}

With the model space defined, equation \ref{Reducedmatrixelement} can be used to obtain double barred matrix elements and thus $B(E1)$ strengths between the states of opposite parity bands. The explicit replacement of all particular labels is shown in equation \ref{ReducedmatrixelementTh} for the transition from excited to ground state band. Due to the lack of effective charges for the model and experimental data of $B(E1)$ strengths, the parameters $a = e_{\pi}\sqrt{\alpha}$ and $b = e_{\nu}\sqrt{1-\alpha}$ are introduced. The process to determine these is explained below.

One interesting feature that emerges for this particular case is that the one-body irreps $(0,1)$ and $(1,2)$ do not couple $(65,6)$ to $(60,8)$, so their contributions are null. The triple barred reduced matrix elements present in this expression require a further calculation as indicated in appendix B. The 9-$(\lambda,\mu)$ symbols denoted by curly brackets as $\{...\}$ and the $SU(3)$ Wigner coefficients $\langle...;...||...\rangle$ are obtained using the libraries in \cite{DYTRYCH2021108137} and the coefficients $\mathcal{V}_{\sigma}^{1}$ come from the tensor expansion of the operator $\mathcal{Q}^{\hspace{2mm}1}_{\sigma\mu}$ shown in table \ref{table1}.

\begin{widetext}
\begin{equation}
\resizebox{0.93\hsize}{!}{$
\begin{aligned} 
\Big\langle\Psi'^{^{224}\text{Th}}_{g.s.b.}(J'^{\Pi=+})\Big|\Big|&
T^{1}(E)
\Big|\Big|\Phi^{^{224}\text{Th}}_{e.b.}
(J^{\Pi=-})\Big\rangle=\\\\ 2.4644\hspace{1mm}\text{fm}&
\Biggl[a\sum_{(\lambda_{\pi_o},\mu_{\pi_o})}\mathcal{V}_{\pi}^{1}(5,6,\lambda_{\pi_o},\mu_{\pi_o},k_{\pi_o})\sum_{ \bar{\rho},\rho_{\pi}}\langle (65,6),0,L; (\lambda_{\pi_o},\mu_{\pi_o}),1||(60,8),0,L' \rangle_{\bar{\rho}}\begin{Bmatrix}
(31,2) & (\lambda_{\pi_o},\mu_{\pi_o}) &  (26,4) &  \rho_{\pi} \\
(34,4) & (0,0) &  (34,4) &  1 \\
(65,6) & (\lambda_{\pi_o},\mu_{\pi_o}) &  (60,8) &  \bar{\rho} \\
1 & 1 & 1 &  \\
\end{Bmatrix}
\\\\& \hspace{30mm}\times\Big\langle 8,\hspace{1mm} [2^4],\hspace{1mm}(26,4)\hspace{1mm}\Big|\Big|\Big|  \Big\{a^{\dagger}_{(5,0)\frac{1}{2}}\times \tilde{a}_{(0,6)\frac{1}{2}}\Big\}^{(\lambda_{\pi_o},\mu_{\pi_o})}\Big|\Big|\Big| 8,[2^3,1]\otimes[1],\hspace{1mm}(25,2)\otimes(6,0),\hspace{1mm}(31,2)\Big\rangle_{\rho_{\pi}}
\\\\&+b\sum_{(\lambda_{\nu_o},\mu_{\nu_o})}\mathcal{V}_{\nu}^{1}(6,7,\lambda_{\nu_o},\mu_{\nu_o},k_{\nu_o})\sum_{\bar{\rho},\rho_{\nu}}\langle (65,6),0,L; (\lambda_{\nu_o},\mu_{\nu_o}),1||(60,8),0,L' \rangle_{\bar{\rho}}\begin{Bmatrix}
(26,4) & (0,0) &  (26,4) &  1 \\
(39,2) & (\lambda_{\nu_o},\mu_{\nu_o}) &  (34,4) &  \rho_{\nu} \\
(65,6) & (\lambda_{\nu_o},\mu_{\nu_o}) &  (60,8) &  \bar{\rho} \\
1 & 1 &  1 &  \\
\end{Bmatrix}
\\\\& \hspace{30mm} \times\Big\langle 8,\hspace{1mm} [2^4],\hspace{1mm}(34,4)\hspace{1mm}\Big|\Big|\Big|  \Big\{a^{\dagger}_{(6,0)\frac{1}{2}}\times \tilde{a}_{(0,7)\frac{1}{2}}\Big\}^{(\lambda_{\nu_o},\mu_{\nu_o})}\Big|\Big|\Big| 8,\hspace{1mm} [2^3,1]\otimes[1],\hspace{1mm}(32,2)\otimes(7,0),\hspace{1mm}(39,2)\Big\rangle_{\rho_{\nu}}\Biggr].
\end{aligned} 
\label{ReducedmatrixelementTh}
$}
\end{equation}
\end{widetext}

The only experimental data available for this isotope involving $B(E1)$ strengths are the ratios $R_J(E1/E2)$
which can be obtained from observed static dipole to quadrupole
ratios $D_0/Q_0$ \cite{ACKERMANN199361, nndc} by the relation
\begin{equation}
\resizebox{0.87\hsize}{!}{$
\begin{aligned} 
R_J(E1/E2)=\frac{B(E1;J\rightarrow J-1)}{B(E2;J\rightarrow J-2)}=\frac{16}{5}\frac{2J-1}{2J-2}\left(\frac{D_0}{Q_0}\right)^2,
\end{aligned} 
\label{BE1estimation}
$}
\end{equation}
its values are shown in table \ref{E1E2Ratios}, third column.

The model only considers transitions from the negative to the positive parity band, however, the contrary process strengths can be figured out given that the modulus squared of double barred reduced matrix element is independent of the direction of the transition. The relation 
\begin{equation}
\resizebox{0.77\hsize}{!}{$
\begin{aligned} 
B(EL;J\rightarrow J') = \Bigg(\frac{2J'+1}{2J+1}\Bigg)^2B(EL;J'\rightarrow J),
\end{aligned} 
\label{BE1downwardsupwards}
$}
\end{equation}
can be derived from equation \ref{BE1}. Notice that the factor $(2J'+1)/(2J+1)$ in equations \ref{BE1} and \ref{BE1downwardsupwards} depends on the convention used for the normalization factor in the Wigner-Eckart theorem. Here we adopted the same convention as the main references \cite{TROLTENIER199553,CASTANOS1987290,RAJU1973433} which is defined in \cite{rose1957elementary}.

As has been stated above, in reference \cite{PhysRevC.101.054306} the $B(E2)$ strengths have been calculated using the equation \cite{SUN2002130} 
\begin{equation}
\resizebox{0.88\hsize}{!}{$
\begin{aligned} 
B(&E2;J\rightarrow J') = \frac{2J'+1}{2J+1}\alpha^2\\&\times\Big|\langle(\lambda,\mu),K,J; (1,1),2||(\lambda,\mu),K,J'\rangle\Big|^2C^{(2)}(\lambda,\mu),
\end{aligned} 
\label{ysun}
$}
\end{equation}
where $C^{(2)}(\lambda,\mu)$ is the $SU(3)$ Casimir operator of order two. The parameter $\alpha^2$ is obtained from the only experimental value of $^{224}$Th which is $B(E2;2_1^{+}\rightarrow 0_1^{+}) = 96 $ (7) W.u. \cite{nndc}. For the $J\rightarrow J-2$ transitions, the parameter has a value of $\alpha^2 = 0.1119$ (82) W.u. Then, by means of equation \ref{ysun}, the intra-band $B(E2; J\rightarrow J-2)$ strengths can be obtained.

\begin{table}[]
\caption{\label{E1E2Ratios}
Ratios of reduced transition probabilities  $R_J(E1/E2)$ and $B(E1)$ strengths for the isotope $^{224}$Th.}
\begin{ruledtabular}
\begin{tabular}{cccc}
\makecell{Transition\\ $J_i^{\Pi_i} \rightarrow J_f^{\Pi_f}$} & 
 \makecell{Th. $B(E1)$ \\  $\times 10^{-2}$ $[\text{W.u.}]$} &
\multicolumn{2}{c}{$R_{J_i}(E1/E2)$} \\
\cmidrule{3-4} & &
\makecell{ Exp. \\  $\times10^{-5}$}  &
\makecell{Th.  \\ $\times10^{-5}$} \\
\hline
& & \\
$1^-\rightarrow 0^+$  & 0.065 & - & -\\
$1^-\rightarrow 2^+$  & 0.470 & - & -\\
$3^-\rightarrow 2^+$  & 0.009 & - & 0.065\\
$5^-\rightarrow 4^+$  & 0.013 & - & 0.081\\
$6^+\rightarrow 5^-$  & 0.634 & 6.38 (14) & 4.224\\
$7^-\rightarrow 6^+$  & 0.118 & - &  0.680\\
$8^+\rightarrow 7^-$  & 0.871 & 5.231 (57) & 5.566\\
$9^-\rightarrow 8^+$  & 0.339 & - & 1.902\\
$10^+\rightarrow 9^-$  & 1.091 & 6.117 (18) & 6.838\\
$11^-\rightarrow 10^+$  & 0.685 & 8.844 (41) & 3.788\\
$12^+\rightarrow 11^-$  & 1.283 & 8.023 (18) & 7.970\\
$13^-\rightarrow 12^+$  & 1.155 & 7.250 (18) & 6.350\\
$14^+\rightarrow 13^-$  & 1.439 & 9.768 (28) & 8.914\\
$15^-\rightarrow 14^+$  & 1.740 & 8.923 (41) & 9.569\\
$16^+\rightarrow 15^-$  & 1.551 & - & 9.638\\
$17^-\rightarrow 16^+$  & 2.424 & 11.22 (19) & 13.385\\
$18^+\rightarrow 17^-$  & 1.619 & - & 10.128\\
\end{tabular}
\end{ruledtabular}
\end{table}

\begin{figure}[t]
\includegraphics[trim={0cm 0 0 0}, clip, width=0.48\textwidth]{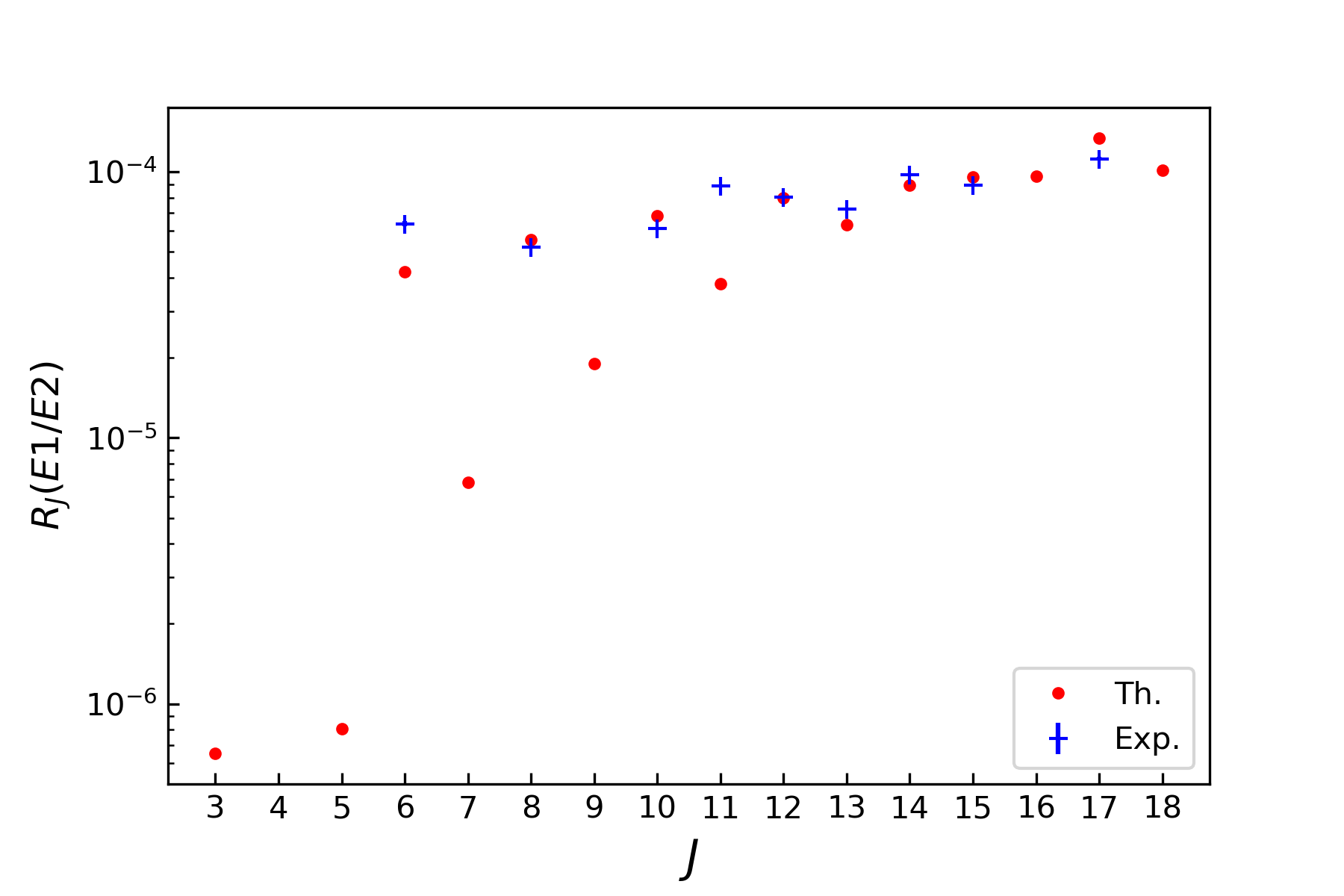}
\caption{\label{RE1E2} Experimental and theoretical ratios $R_{J}(E1/E2)$.} 
\end{figure}
\begin{figure}[t]
\includegraphics[trim={1.5cm 0 0 0}, clip, width=0.48\textwidth]{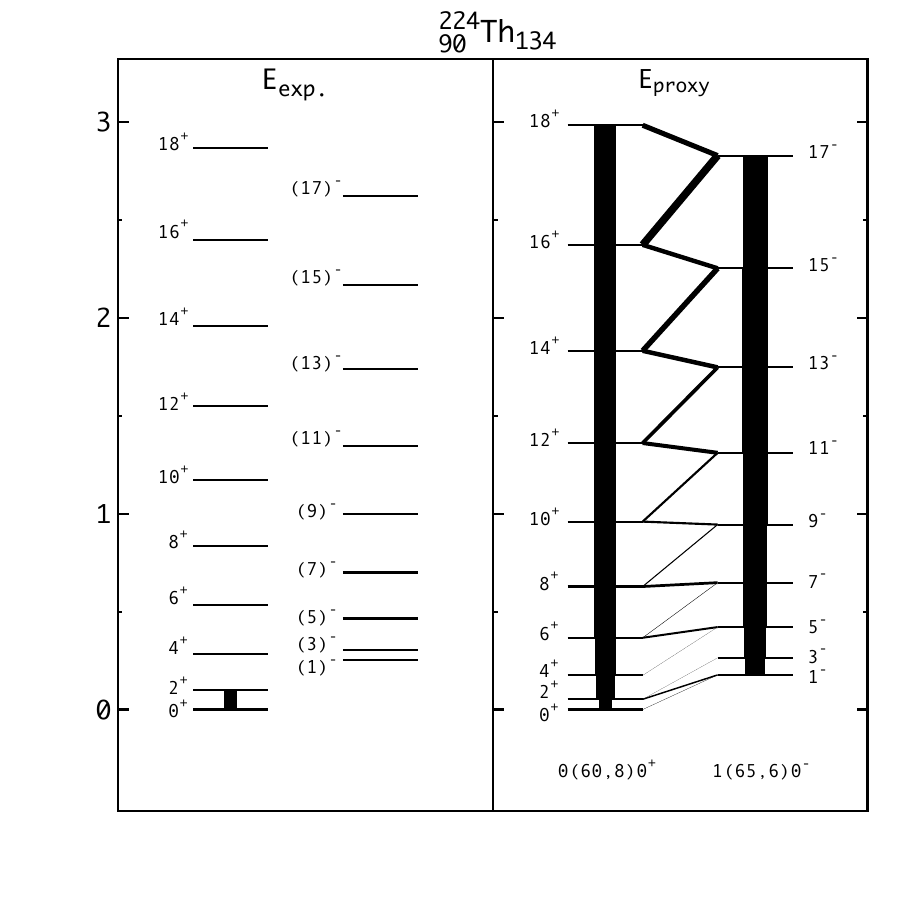}
\caption{\label{scheme} Experimental and theoretical low energy spectra of $^{224}$Th. The energy is measured
in MeV. Electromagnetic $B(E2)$ intra-band and $B(E1)$ inter-band transition strengths are shown as solid lines between states where the width is proportional to the numerical value. The labels of the theoretical section are $n(\lambda,\mu)K^{\Pi}$.} 
\end{figure}
Using this result, the ratios $R_J(E1/E2)$ can be formulated as a function of parameters $a$ and $b$ from the quotient of equations \ref{BE1} and \ref{ysun} using reduced matrix elements in \ref{ReducedmatrixelementTh}. The best fit of the function $R_J(E1/E2, a,b)$ to experimental data delivers the values of $a = 2.27e $ (14) and  $b = -1.18e$ (10), where $e$ is the elementary charge and the uncertainties are  twice the standard deviations obtained by the fitting procedure.

The theoretical $B(E1)$ values can be obtained by evaluating \ref{BE1} with matrix elements \ref{ReducedmatrixelementTh} and the parameters obtained in the last paragraph. Its values are shown in table \ref{E1E2Ratios} second column along with $R_J(E1/E2)$ ratios also in figure \ref{RE1E2}. In figure \ref{scheme} is shown the theoretical scheme obtained previously in \cite{PhysRevC.101.054306} with the $B(E1)$ values from this work included.  

\begin{figure}[t]
\includegraphics[trim={0cm 0 0 0}, clip, width=0.48\textwidth]{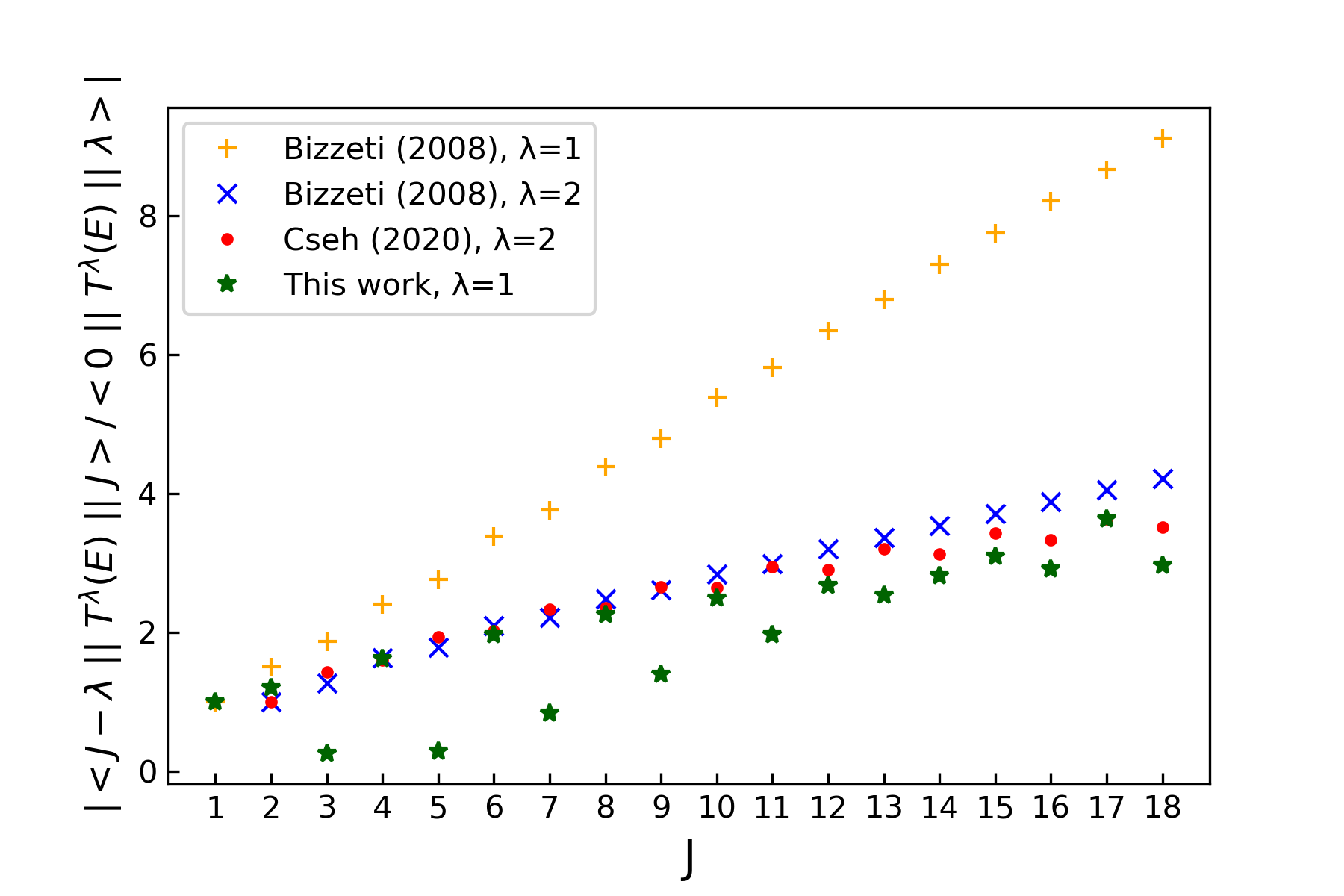}
\caption{\label{ME_Ratios} Normalized matrix elements for dipole and quadrupole transitions of $^{224}$Th from references \cite{PhysRevC.101.054306}, \cite{PhysRevC.77.024320} and this work.} 
\end{figure}

Reduced matrix elements for the intra-band and inter-band transitions of $^{224}$Th  have been studied previously in \cite{PhysRevC.77.024320} under a collective nuclear model \cite{PhysRevC.70.064319}. Due to parameter dependence of these, a normalization with respect to the lowest transition is required. In figure \ref{ME_Ratios} are shown the normalized matrix elements for both dipole and quadrupole transitions for their collective model and the symmetry-based model of this work. The differences between \cite{PhysRevC.77.024320} and this work for the matrix elements of the dipole transitions emerge from the convention in the normalization factor of the Wigner-Eckart theorem, but physical observables converge to close values.

\section{CONCLUSIONS}

The mathematical form of the electric dipole single-particle transition operator $T^1_{\mu}(E)$ was obtained for the actinide region as a sum of irreducible tensors of $SU(3)$. The reduced matrix elements of this operator were obtained in a general form involving recoupling coefficients and two-shell one-body $SU(3)$ reduced  matrix elements. This later factors cannot be computed directly with already developed libraries, thus, additional calculations are needed as explained in appendix B, where a reduction to single-shell matrix elements products is performed. This methodology and the computer codes developed can be extended to other nuclear regions studied under symmetry-based models where dipole transitions appear. The theoretical modeling of this work embedded in the SAQM carries three parameters, namely the effective neutron and proton charges $e_{\pi}$, $e_{\nu}$ and the superposition parameter $\alpha$ for the contributions of each nucleon type (proton or neutron) to the total excited nuclear wavefunction.

These results were applied to the isotope $^{224}$Th to obtain ratios $R_J(E1/E2)$ and $B(E1)$ strengths. Due to the lack of experimental $B(E1)$ data and effective charges for this model, the parameters where condensed in $a = e_{\pi}\sqrt{\alpha}$ and $b = e_{\nu}\sqrt{1-\alpha}$. The numerical data obtained were  $a = 2.27e $ and  $b = -1.18e$, which indicate larger values of effective charges for the actinide region than other $SU(3)$-based  models \cite{TROLTENIER199553}. This may be connected to the larger dimension of the irreps in the proxy-$SU(3)$ scheme due to the minimum truncation of the complete model space. More research is required in this matter.

The $B(E1)$ values obtained lie mostly in the order of $10^{-2}-10^{-3}$ W.u., which according to \cite{OctupoleButler} is an indicator of octupole deformation. The $B(E1)$ value of the transition $3^-\rightarrow 2^+$ is predicted to be two orders of magnitude lower from the algebraic structure, but can also be attributed to the fact that no $D_0/Q_0$ data is available to that transition making it an extrapolation of the model. In comparison, the values obtained in  \cite{PhysRevC.77.024320} follow a soft pattern while in this symmetry-based model a marked differentiation between the bands is observed. The deviation between the reduced matrix elements are due to the normalization convention in the Wigner-Eckart theorem which affect the fitting of $R_J(E1/E2)$. Despite these differences, physical observables converge to similar values, as they must if consistency is maintained.

The lack of experimental data is challenging for the theoretical models in this nuclear region. We hope to motivate research towards this direction from the experimental side as well. These quartet nuclei happen to be connected through a chain of alpha decays which adds extra interest to these isotopes. The results obtained in this work can be extended directly to other actinides and heavy nuclei. Currently, work in the inclusion of octupole residual interaction is in progress.


\begin{acknowledgments}

We would like to thank professors J. G. Hirsch, V. K. B. Kota, T. Dytrych, V. G. Gueorguiev, F. Pan, D. Langr, K. P. Drumev, C. Bahri, D. Jenkins, J. Cseh and B. Pritychenko  for very valuable discussions and advice. We thank the Institute of Physics at University of Antioquia for supporting this research. A warm welcome to the world to Martín Valencia Salazar.

\end{acknowledgments}


\appendix

\section{One-body spherical tensor expansion in $SU(3)$}

A spherical tensor can be expanded in angular momentum second quantization by well known methods \cite{talmi1993simple}. Each oscillator shell $\eta$ has a closed $SU(3)$ algebra which is constructed by the operators $a^{\dagger}_{(\eta,0),l,m_l,\frac{1}{2},m_s}$ and $\tilde{a}_{(0,\eta),l,m_{l},\frac{1}{2},m_{s}} = (-1)^{\eta+l-m_l+\frac{1}{2}-m_s}{a}_{(\eta,0),l,m_{l},\frac{1}{2},m_{s}}$. The upper algebra labels $[1],(\eta,0)$ and  $[1],(0,\eta)$ are assigned to $a^{\dagger}_{(\eta,0),l,m_{l},\frac{1}{2},m_{s}}$ and $\tilde{a}_{(0,\eta),l,m_{l},\frac{1}{2},m_{s}}$ respectively by physical arguments so that they transform as $SU(3)$ tensors. Coupling the $SU(3)$ irreps results in the expansion of the spherical tensor into a sum over $(\lambda_o,\mu_o)$ of $SU(3)$ tensors.

A one body-operator conserves the number of particles, so it will involve one construction and one destruction operators coupled to total irreps according to the algebra chain $\Big\{SU(3)\supset SO_L(3)\Big\}\otimes SU_S(2) \supset SU_J(2)\supset U_{M_J}(1)$ for the spatial, spin and total angular momentum quantum numbers. For a more general case involving several shells, a sumation over oscillator shells must be included where selection rules over them will emerge. The expansion of a general operator can be seen in equation \ref{operatorgeneral}. Details of the derivation of such equation can be provided under request.

\begin{widetext}
\begin{equation}
\resizebox{0.82\hsize}{!}{$
\begin{aligned} 
O^{\hspace{3mm} L_o S_o;J_o}_{\sigma\hspace{8mm}M_{J_o}} =&\sum_{(\lambda_o,\mu_o),K_o} \sum_{\eta,\eta'}\sum_{l,l'}(-1)^{\eta}\sqrt{\frac{2(2l'+1)}{(2L_o+1)(2S_o+1)}}\Big\langle \eta',l',\frac{1}{2} \Big|\Big| O^{\hspace{3mm} L_o \hspace{0mm}S_o}_{\sigma} \Big|\Big| \eta,l,\frac{1}{2}\Big\rangle\\&\times\langle (\eta',0),0,l'; (0,\eta),0,l||(\lambda_o,\mu_o),K_o,L_o \rangle_{\rho_o=1}\Big\{a^{\dagger}_{(\eta',0)\frac{1}{2}}\times \tilde{a}_{(0,\eta)\frac{1}{2}}\Big\}^{\rho_o=1(\lambda_o,\mu_o)K_oL_oS_o;J_o}_{\hspace{25mm}M_{J_o}}\\&=\sum_{(\lambda_o,\mu_o),K_o}\sum_{\eta,\eta'}\mathcal{V}_{\sigma}(\eta',\eta,\lambda_o,\mu_o, K_o)\Big\{a^{\dagger}_{(\eta',0)\frac{1}{2}}\times \tilde{a}_{(0,\eta)\frac{1}{2}}\Big\}^{\rho_o=1(\lambda_o,\mu_o)K_oL_oS_o;J_o}_{\hspace{25mm}M_{J_o}}.
\end{aligned} 
\label{operatorgeneral}
$}
\end{equation}
\end{widetext}

\section{Multi-shell reduced matrix element computation}

The multi-shell one-body matrix elements that emerge in this work require a recoupling between the operator and state labels to simplify them into products of single-shell matrix elements. This process enables straightforward computation using existing libraries. Since each shell has an $SU(3)$ symmetry, both spaces can be separated and calculated independently carrying recoupling coefficients in such process. The resulting expression can be found in equation \ref{recouplingOB} where $\chi\{...\}$ represents the unitary $9-j$ symbols and the identity \ref{atildetoadagger} was used. Details of the derivation of such equations can be provided under request. The 9-$(\lambda,\mu)$ and $\langle...|||a^{\dagger}_{...}|||...\rangle$ values can be computed straightforwardly with libraries \cite{BAHRI199459,DYTRYCH2021108137}.

\begin{widetext}
\begin{equation}
\resizebox{0.93\hsize}{!}{$
\begin{aligned} 
\Big\langle[f'_-]\otimes[f'_+],(\lambda'_-\mu'_-)\otimes(\lambda'_+\mu'_+),\rho'(\lambda'\mu'),S'\Big|\Big|\Big| \big\{a^{\dagger}_{(\eta'0)\frac{1}{2}}\tilde{a}_{(0\eta)\frac{1}{2}}\big\}^{\rho_0=1,(\lambda_0\mu_0)S_0}\Big|\Big|\Big|[f_-]\otimes[f_+],(\lambda_-\mu_-)\otimes(\lambda_+\mu_+),\rho(\lambda\mu),S\Big\rangle_{\bar{\rho}}
\\= \begin{Bmatrix}
  (\lambda_-,\mu_-) & (\eta',0) & (\lambda'_-,\mu'_-) & \rho=1\\
  (\lambda_+,\mu_+) & (0,\eta) & (\lambda'_+,\mu'_+) & \rho=1\\
  (\lambda,\mu) & (\lambda_o,\mu_o) & (\lambda',\mu') & \bar{\rho}\\
  \rho & \rho=1 & \rho'&
 \end{Bmatrix}
\chi\begin{Bmatrix}
  S_- & \frac{1}{2} & S'_- \\
  S_+ & \frac{1}{2} & S'_+ \\
  S & S_o & S' \\
 \end{Bmatrix}(-1)^{\eta+\frac{1}{2}+S_+-S'_++\lambda'_+-\lambda_++\mu'_+-\mu_+}\sqrt{\frac{\text{dim}(\lambda_+\mu_+)}{\text{dim}(\lambda'_+\mu'_+)}}\frac{\sqrt{2S_++1}}{\sqrt{2S'_++1}}
\\\\ \times\Big\langle[f'_-],(\lambda'_-\mu'_-),S'_-\Big|\Big|\Big|a^{\dagger}_{(\eta'0)\frac{1}{2}}\Big|\Big|\Big| [f_-],(\lambda_-\mu_-)S_-\Big\rangle_{\rho=1}
\Big\langle[f_+],(\lambda_+\mu_+),S_+\Big|\Big|\Big|a^{\dagger}_{(\eta0)\frac{1}{2}}\Big|\Big|\Big|[f'_+],(\lambda'_+\mu'_+),S'_+\Big\rangle_{\rho=1},
\end{aligned} 
\label{recouplingOB}
$}
\end{equation}

\begin{equation}
\resizebox{0.93\hsize}{!}{$
\begin{aligned} 
\Big\langle[f''](\lambda''\mu''),S''\Big|\Big|\Big|\tilde{a}_{(0\eta)\frac{1}{2}}\Big|\Big|\Big|[f](\lambda\mu),S\Big\rangle_{\rho=1}
= (-1)^{\eta+\frac{1}{2}+S-S''+\lambda''-\lambda+\mu''-\mu}\sqrt{\frac{\text{dim}(\lambda\mu)}{\text{dim}(\lambda''\mu'')}}\sqrt{\frac{2S+1}{2S''+1}}\Big\langle[f](\lambda\mu),S\Big|\Big|\Big|a^{\dagger}_{(\eta0)\frac{1}{2}}\Big|\Big|\Big|[f''](\lambda''\mu''),S''\Big\rangle_{\rho=1}.
\end{aligned} 
\label{atildetoadagger}
$}
\end{equation}
\end{widetext}


\nocite{*}

\bibliography{bib}

\end{document}